\documentclass[useAMS,usenatbib]{mn2e}

\usepackage{latexsym,graphicx,natbib}

\usepackage{color}

\newcommand\kms{{\rm\,km\,s^{-1}}}

\newcommand{\MC}{\multicolumn}

\def\apgt{\ {\raise-.5ex\hbox{$\buildrel>\over\sim$}}\ }
\def\aplt{\ {\raise-.5ex\hbox{$\buildrel<\over\sim$}}\ }

\title[WS1: bona fide luminous blue variable]{WS1: one more new Galactic bona fide luminous blue variable\footnotemark[0]\thanks{
{Based on observations obtained with the Southern African Large
Large Telescope (SALT), programmes \mbox{2010-1-RSA\_OTH-001},
\mbox{2013-1-RSA\_OTH-014} and \mbox{2013-2-RSA\_OTH-003}.}}}
\author[A.Y Kniazev et al.]
       {A. Y.~Kniazev,$^{1,2,3}$\thanks{E-mail: akniazev@saao.ac.za}
    V. V.~Gvaramadze,$^{3,4,5}$
    L. N.~Berdnikov$^{6,3,5}$ \\
    $^{1}$South African Astronomical Observatory, PO Box 9, 7935 Observatory, Cape Town, South Africa \\
    $^{2}$Southern African Large Telescope Foundation, PO Box 9, 7935 Observatory, Cape Town, South Africa \\
    $^{3}$Sternberg Astronomical Institute, Lomonosov Moscow State University, Universitetskij Pr. 13, Moscow 119992, Russia\\
    $^{4}$Space Research Institute, Russian Academy of Sciences, Profsoyuznaya 84/32, Moscow 117997, Russia \\
    $^{5}$Isaac Newton Institute of Chile, Moscow Branch, Universitetskij Pr. 13, Moscow 119992, Russia \\
    $^{6}$Astronomy and Astrophysics Research division,  Entoto Observatory and Research Center, P.O.Box 8412, Addis Ababa, Ethiopia \\
    }
\begin{document}

\date{Accepted 2015 February 3. Received 2015 February 3; in original form 2014 November 24}


\maketitle

\label{firstpage}

\begin{abstract}
In this Letter, we report the results of spectroscopic and
photometric monitoring of the candidate luminous blue variable
(LBV) WS1, which was discovered in 2011 through the detection of a
mid-infrared circular shell and follow-up optical spectroscopy of
its central star. Our monitoring showed that WS1 brightened in the
$B$, $V$ and $I$ bands by more than 1 mag during the last three
years, while its spectrum revealed dramatic changes during the
same time period, indicating that the star became much cooler. The
light curve of WS1 demonstrates that the brightness of this star
has reached maximum in 2013 December and then starts to decline.
These findings unambiguously proved the LBV nature of WS1 and
added one more member to the class of Galactic bona fide LBVs,
bringing their number to sixteen (an updated census of these
objects is provided).
\end{abstract}

\begin{keywords}
line: identification -- stars: emission-line, Be -- stars:
evolution -- stars: individual: [GKM2012]\,WS1 -- stars: massive.
\end{keywords}

\section{Introduction}
\label{sec:intro}

Luminous blue variables (LBVs) are hot luminous supergiant stars
showing strong photometric and spectral variability on time-scales
from years to decades (Humphreys \& Davidson 1994; van Genderen
2001). The origin of this variability is still poorly understood,
partly because of the small number (about a dozen) of confirmed
members of this class of massive evolved stars. It is believed
that during the LBV phase, the massive stars lose a significant
fraction of their mass, which play an important role in their
subsequent evolution (Langer et al. 1994) and is manifested in the
presence of circumstellar nebulae around the majority of LBVs
(Nota et al. 1995; Clark, Larionov \& Arkharov 2005). Detection of
nebulae reminiscent of those associated with LBVs and related
stars provides a powerful tool for revealing evolved massive stars
(Waters et al. 1996; Egan et al. 2002; Clark et al. 2003;
Gvaramadze et al. 2009, Gvaramadze, Kniazev \& Fabrika 2010c;
Gvaramadze et al. 2010a; Wachter et al. 2010; Burgemeister et al.
2013). There are also indications that LBVs might be immediate
progenitors of supernovae (Kotak \& Vink 2006; Smith et al. 2007;
Groh, Meynet \& Ekstr\"{o}m 2013). This makes them particularly
interesting objects to study because some of the already known
LBVs could soon end their lives in spectacular explosions.
Disclosing of new members of this class of evolved massive stars
is therefore of high importance.

Recent searches for LBVs using follow-up spectroscopy of candidate
evolved massive stars, revealed through detection of their
mid-infrared circumstellar nebulae with the {\it Spitzer Space
Telescope} and {\it Wide-field Infrared Survey Explorer} ({\it
WISE}), resulted in nearly doubling the number of stars with
spectra typical of LBVs (Gvaramadze et al. 2010a,b, 2012, 2014;
Wachter et al. 2010, 2011; Stringfellow et al. 2012a,b; Flagey et
al. 2014). However, only few of these stars show variability
inherent to bona fide LBVs (e.g. Gvaramadze et al. 2014), while
others are in a quiescent state since their discovery and are
considered as candidate LBVs (cLBVs). It is worthy therefore to
monitor the known cLBVs to search for their expected spectral and
photometric variability.

In this Letter, we report the results of spectroscopic and
photometric monitoring of WS1 -- one of the two cLBVs discovered
in Gvaramadze et al. (2012; Paper\,I hereafter) through the
detection of circular shells with {\it WISE} and follow-up
spectroscopy of their central stars with the Southern African
Large Telescope (SALT). Using archival and contemporary
photometry, we found in Paper\,I that WS1\footnote{Named in the
SIMBAD data base as [GKM2012]\,WS1.} brightened in the $R$ and $I$
bands by 0.68$\pm$0.10 mag and 0.61$\pm$0.04 mag, respectively,
during 13--18 yr. This prompted us to continue observations of WS1
(Section\,\ref{sec:obs}), which unambiguously proved the LBV
nature of WS1 (Section\,\ref{sec:LBV}) and added one more member
to the class of Galactic bona fide LBVs, bringing their number to
sixteen (see Section\,\ref{sec:con} for the current census of
these stars).

\begin{figure*}
\begin{center}
\includegraphics[angle=270,width=0.8\textwidth,clip=]{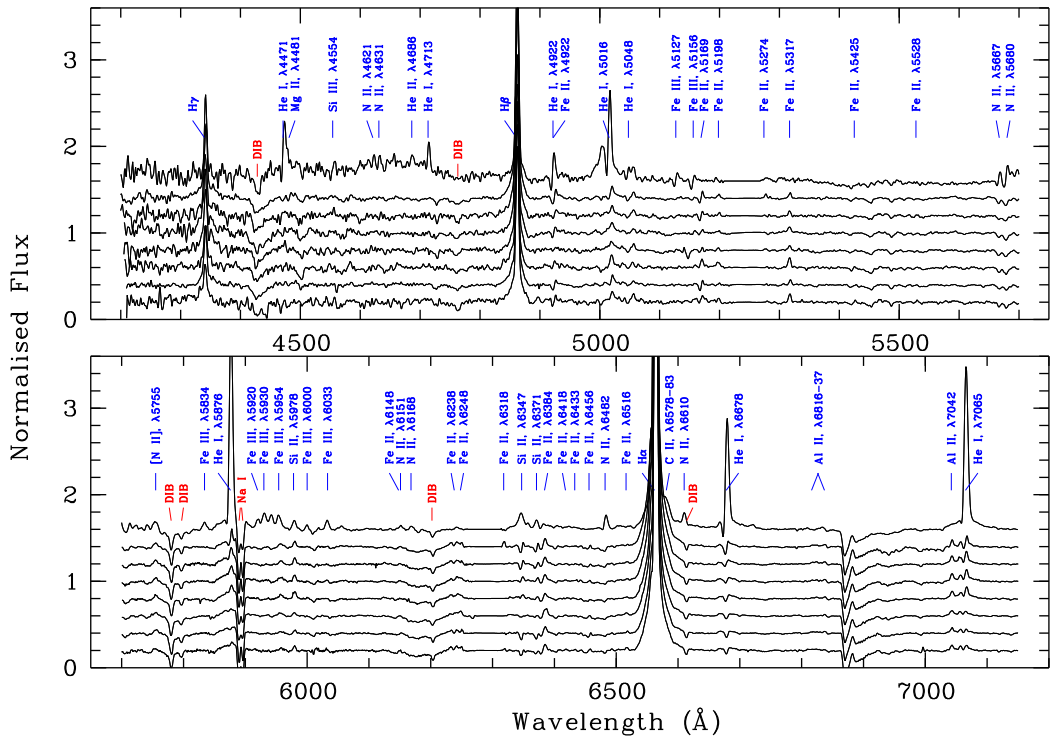}
\end{center}
\caption{Evolution of the (normalized) spectrum of WS1 since its
discovery in 2011 (from top to bottom; see Table\,\ref{tab:Obs}
for the dates of the observations). Principal lines and most
prominent diffuse interstellar bands (DIBs) are indicated. For
clarity, the spectra are offset by 0.2 continuum flux unit.}
\label{fig:spec}
\end{figure*}

\section{Observations of WS1}
\label{sec:obs}

\subsection{SALT spectroscopy}
\label{sec:SALT}

The first spectrum of WS1 was obtained with the SALT (Buckley,
Swart \& Meiring 2006; O'Donoghue et al. 2006) on 2011 June 12
with the Robert Stobie Spectrograph (Burgh et al. 2003; Kobulnicky
et al. 2003) in the long-slit mode. This spectrum was presented
for the first time and discussed in detail in Paper\,I. It shows
strong H and He\,{\sc i} emission lines, numerous prominent metal
lines of N\,{\sc ii}, Fe\,{\sc iii} and Si\,{\sc ii} in emission,
and the weak He\,{\sc ii} $\lambda 4686$ emission (see figs\,3$-$5
in Paper\,I and Fig.\,\ref{fig:spec}), which are typical of hot
LBVs during the visual minimum (e.g. Stahl et al. 2001). Based on
this spectrum and photometric variability (see
Section\,\ref{sec:intro}), and the presence of a mid-infrared
shell around WS1 (see fig.\,1 in Paper\,I), we classified this
star as a cLBV.

In the course of spectral monitoring of WS1 in 2013--2014, we
obtained seven more spectra with the SALT. The PG900 grating was
used for these observations. In all cases, the spectral range of
$4200-7300$~\AA\ was covered with a final reciprocal dispersion of
$0.97$~\AA\ pixel$^{-1}$. The spectral resolution full width at
half-maximum (FWHM) was of $5.05\pm$0.15~\AA. In most cases, one
short exposure was obtained additionally to get non-saturated
H$\alpha$ line. A Xe lamp arc spectrum was taken immediately after
all science frames. Spectrophotometric standard stars were
observed during twilight time for relative flux calibration.
Absolute flux calibration is not feasible with SALT because the
unfilled entrance pupil of the telescope moves during the
observations. The log of all our spectroscopic observations of WS1
is listed in Table\,\ref{tab:Obs}.

Primary reduction of the data was done in the standard way with
the SALT science pipeline (Crawford et al. 2010). Following
long-slit data reduction was carried out in the way described in
Kniazev et al. (2008).

\begin{table}
\caption{Journal of the observations.} \label{tab:Obs}
\begin{tabular}{llcc} \hline
Date & Exposure & Slit/Seeing & JD \\
& (min) & (arcsec) & (d) \\
 \hline
 2011 June 12$^{a}$  & 3$\times$10             &  1.00/2.1    &  2455725  \\
 2013 April 22       & 1$\times$1,1$\times$10  &  1.25/1.0    &  2456405  \\
 2013 May 23         & 1$\times$1,1$\times$10  &  1.25/2.0   &  2456436  \\
 2013 June 18        & 1$\times$0.5,1$\times$5 &  1.25/2.5    &  2456462  \\
 2013 July 22        & 1$\times$1,1$\times$10  &  1.25/2.5    &  2456496  \\
 2014 January 13     & 2$\times$1,1$\times$13  &  1.25/3.5    &  2456671  \\
 2014 March 13       & 1$\times$1,1$\times$10  &  1.25/1.5    &  2456730  \\
 2014 April 18       & 1$\times$10             &  1.25/1.6    &  2456766  \\
 \hline
   \MC{4}{l}{$^{a}$Presented for the first time in Paper\,I.}
\end{tabular}
\end{table}

\subsection{Photometry}
\label{sec:phot}

To search for photometric variability of WS1, we occasionally
obtained its CCD photometry with the 76-cm telescope of the South
African Astronomical Observatory during our observing runs in
2011--2014. We used an SBIG ST-10XME CCD camera equipped with
$BVI_{\rm c}$ filters of the Kron-Cousins system (for details see
Berdnikov et al. 2012). The results are presented in
Table\,\ref{tab:phot}. To this table we also added archival
photometry from Paper\,I, which is based on the data from the
USNO\,B-1 catalogue (Monet et al. 2003), and photometry derived
from two unsaturated $V$ and $I$ band acquisition images obtained
with the SALT and calibrated using the secondary photometric
standards established from the 76-cm telescope data.

\begin{table*}
  \caption{Archival and contemporary photometry of WS1. The measurements
  were obtained with the 76-cm telescope, unless otherwise is noted.}
  \label{tab:phot}
    \begin{tabular}{lcccl}
      \hline
      Date                    & $B$             & $V$            & $I$            & JD \\
      \hline
      1976 March 8$^{a}$      & 17.50$\pm$0.10  & --             & --             & 2442846 \\
      1979 June 7$^{a}$       & --              & --             & 12.60$\pm$0.10 & 2444032 \\
      2011 June 12$^{b}$      & --              & 15.25$\pm$0.04 & --             & 2455725 \\
      2011 September 23$^{b}$ & 17.36$\pm$0.03  & 15.31$\pm$0.03 & 12.18$\pm$0.03 & 2455828 \\
      2011 December 10        & 17.05$\pm$0.06  & 14.93$\pm$0.03 & 11.94$\pm$0.01 & 2455905 \\
      2012 May 6              & 16.61$\pm$0.03  & 14.50$\pm$0.01 & 11.36$\pm$0.01 & 2456053 \\
      2013 January 13         & 16.23$\pm$0.03  & 14.06$\pm$0.01 & 10.84$\pm$0.01 & 2456305 \\
      2013 May 23$^{c}$       & --              & 13.98$\pm$0.03 & --             & 2456436 \\
      2013 December 31        & 16.05$\pm$0.03  & 13.85$\pm$0.01 & 10.74$\pm$0.01 & 2456657 \\
      2014 January 13         & 16.04$\pm$0.03  & 13.91$\pm$0.01 & 10.74$\pm$0.01 & 2456670 \\
      2014 January 19         & 16.06$\pm$0.03  & 13.90$\pm$0.01 & 10.78$\pm$0.01 & 2456676 \\
      2014 January 26         & 16.05$\pm$0.03  & 13.90$\pm$0.01 & 10.76$\pm$0.01 & 2456683 \\
      2014 March 31           & 16.07$\pm$0.03  & 13.86$\pm$0.02 & 10.77$\pm$0.02 & 2456748 \\
      2014 April 9            & 16.14$\pm$0.03  & 13.92$\pm$0.02 & 10.82$\pm$0.02 & 2456757 \\
      2014 April 16           & 16.23$\pm$0.03  & 13.96$\pm$0.02 & 10.81$\pm$0.02 & 2456764 \\
      2014 April 18$^{c}$     & --              & --             & 10.78$\pm$0.04 & 2456766 \\
      2014 May 12             & 16.20$\pm$0.03  & 13.99$\pm$0.02 & 10.86$\pm$0.02 & 2456790 \\
      \hline
  \MC{5}{l}{$^{a}$USNO\,B-1; $^{b}$Paper\,I; $^{c}$SALT.}
    \end{tabular}
    \end{table*}

\section{WS\,1: a bona fide LBV}
\label{sec:LBV}

Fig.~\ref{fig:spec} presents a montage of all eight (normalized)
spectra of WS1 with the date of the observations (listed in
Table\,\ref{tab:Obs}) increasing from top to bottom. Comparison of
the 2011's spectrum with those obtained in 2013--2014 revealed
dramatic changes in its appearance. One can see that already in
the second spectrum (2013 April 22) the He\,{\sc i} emission lines
are almost disappeared. Similarly, Fe\,{\sc iii} lines are
disappeared as well, while numerous Fe\,{\sc ii} emissions became
prominent. Besides, most of the N\,{\sc ii} and Si\,{\sc ii} lines
changed from emission in 2011 into almost pure absorption in the
later spectra. All these changes are typical of bona fide LBVs
(e.g. Stahl et al. 2001, Groh et al. 2009) and indicate that the
effective temperature, $T_{\rm eff}$, of WS1 has decreased
drastically during the last one and half years.

In Paper\,I, we noted that the spectrum of WS1 is very similar to
that of the bona fide LBV AG Car during the epoch of a minimum in
1985$-$90, when it was a WN11 star (Stahl et al. 2001) with
$T_{\rm eff}$$\approx$22\,000 K (Groh et al. 2009), and argued
that WS1 could be classified as WN11 as well. Using the Stellar
Spectral Flux Library by Pickles (1998), we also found the colour
excess towards WS1 of $E(B-V)=2.40$ mag by matching the dereddened
spectral slope of this star with those of stars of similar $T_{\rm
eff}$. Applying the same $E(B-V)$ to deredden the 2013--2014
spectra of WS1, we found that its $T_{\rm eff}$ decreased to
$\approx$12\,000 K by 2013 April 22 and remained around this value
ever since.

With the above temperature estimates, one can attempt to check
whether the spectrophotometric variability of WS1 is accompanied
by a change in the bolometric luminosity (cf. Clark et al. 2009;
Groh et al. 2009; Maryeva \& Abolmasov 2012). Using the $V$
magnitudes from Table\,\ref{tab:phot} and adopting the visual
bolometric corrections of $\approx$$-2.4$ and $-$0.8 mag,
respectively, for the hot and cool states of WS1 (cf. Groh et al.
2009; Crowther, Lennon \& Walborn 2006), one finds that the
bolometric luminosity has decreased by a factor of 1.2. Whether
this modest decrease is real or simply a result of inaccuracy of
our estimate could be proved with a detailed spectral analysis of
WS1, which is, however, beyond the scope of this Letter and will
be presented elsewhere.

\begin{figure}
\begin{center}
\includegraphics[angle=-90,width=0.8\columnwidth,clip=]{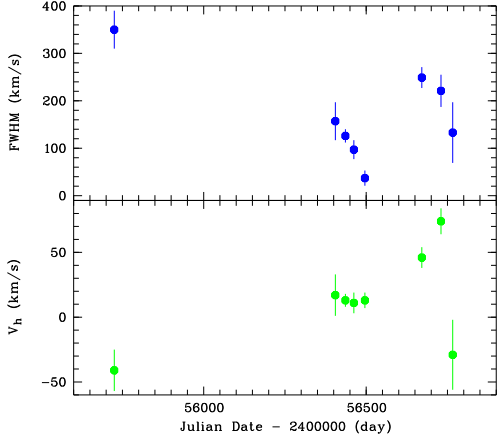}
\end{center}
\caption{Changes of the FWHM and $v_{\rm h}$ of the [N\,{\sc ii}]
$\lambda 5755$ line with time (1$\sigma$ errors are indicated by
vertical bars).} \label{fig:5755}
\end{figure}

\begin{figure}
\begin{center}
\includegraphics[angle=-90,width=0.8\columnwidth,clip=]{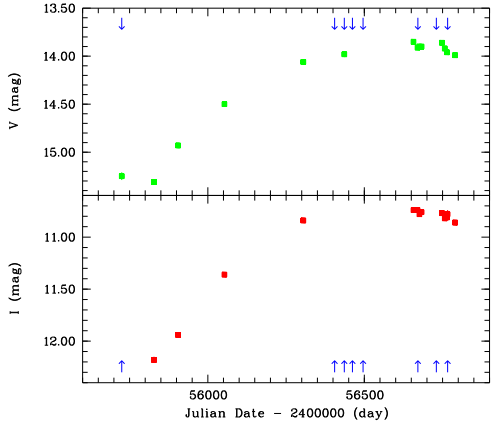}
\end{center}
\caption{Light curves of WS1 in the $V$ and $I$ bands since its
discovery in 2011. 1$\sigma$ error bars are indicated, but in most
cases they are within the size of the data points (boxes). The
arrows mark the dates of the SALT spectra.} \label{fig:phot}
\end{figure}

The only forbidden line visible in the spectra, [N\,{\sc ii}]
$\lambda 5755$, is also weakened considerably. Its FWHM could be
used to derive the stellar wind velocity, $v_\infty$ (cf.
Leitherer et al. 1985). Fig.\,\ref{fig:5755} (upper panel) shows
evolution of $v_\infty$ from spectrum to spectrum. As expected,
the wind velocity slows down significantly with the decrease of
$T_{\rm eff}$, reaching its minimum value of $\approx$40$\pm$20$
\, \kms$ near the visual maximum of WS1, then it increased by
almost six times during about a half year and then slows down
again by a factor of 2 during the next three months along with the
brightness decline of the star (cf. Fig\,\ref{fig:phot}). It is
likely that these changes in $v_\infty$ reflect changes in the
escape velocity from the stellar surface, which is inversely
proportional to the stellar radius, $R_*$, and therefore should be
minimum at the visual maximum of the star when $R_*$ reaches its
maximum value (e.g. Stahl et al. 2001; Leitherer et al. 1994; Groh
et al. 2009). It is worth noting that similar changes in
$v_\infty$ were revealed in AG\,Car, which decreased from 250 to
50 $\kms$ on a time-scale of about one year (Leitherer et al.
1994).

It is also believed that the heliocentric radial velocity, $v_{\rm
h}$, of the [N\,{\sc ii}] $\lambda 5755$ line could be used to
estimate the systemic velocity of the star (Stahl et al. 2001).
For WS1 this method, however, does not work because $v_{\rm h}$ of
the [N\,{\sc ii}] line varies from spectrum to spectrum (see the
lower panel of Fig.\,\ref{fig:5755}). This variability may be
caused by changes in the wind velocity.

Fig.~\ref{fig:phot} plots $V$ and $I$ band light curves of WS1
since the discovery of this star in 2011 with arrows indicating
times when the spectra were obtained with the SALT. For each data
point (square) we give 1$\sigma$ error bars, which in most cases
are less than the size of the squares themselves and therefore
they cannot be discerned. One can infer from Fig.~\ref{fig:phot}
and Table~\ref{tab:phot} that WS1 brightened almost smoothly in
the $B$, $V$ and $I$ bands until the end of 2013 with the net
brightness increase of $\approx$1.3 mag in the $B$ band and
$\approx$1.5 mag in the other two ones. Since then the brightness
of the star declines. Thanks to the good time cadence of
observations in 2014, one can see that the brightness decrease is
not smooth, but is accompanied by short-term small-amplitude
variability. Correspondingly, one cannot exclude the possibility
that the overall brightening of WS1 in 2011--2013 was accompanied
by the similar (or higher amplitude) variability as well.

The brightness increase of WS1 would be even higher if one takes
into account the archival photometry. From Table\,\ref{tab:phot}
it follows that WS1 has brightened in the $B$ and $I$ bands by
$\approx$1.5 and 1.9 mag, respectively, in last $\approx$40 yr. We
caution, however, that during this time interval WS1 may well
experience several S Dor-like outbursts and that the archival
photometry might correspond to one of the brightness minima on the
light curve of this star.

To conclude, our photometric and spectroscopic observations
indicate that currently WS1 experiences an S\,Dor-like outburst
and that this star has already passed through the maximum of its
visual brightness.

\section{Galactic bona fide LBVs}
\label{sec:con}

\begin{table*}
\caption{Census of the Galactic bona fide LBVs. The LBVs detected
after a similar census by Clark et al. (2005) was published are
marked by bold face. The objects with detected circumstellar
nebulae are starred.} \label{tab:LBV}
\begin{tabular}{p{2.7cm}p{4.8cm}p{4.3cm}p{3.3cm}}
\hline HR\,Car$^\ast$ & $\eta$\,Car$^\ast$ & AG\,Car$^\ast$ & Wray\,15-751$^\ast$ (Wra\,751$^\ast$)\\
{\bf $[$GKM2012$]$\,WS1}$^\ast$ & {\bf Wray\,16-137}$^\ast$ & Cl*\,Westerlund\,1\,W\,243 (W243) & HD\,160529 \\
GCIRS\,34W & {\bf [MMC2010]\,LBV\,G0.120-0.048}$^\ast$ & qF\,362 (FMM\,362) & HD\,168607 \\
{\bf MWC\,930}$^\ast$ & G24.73+0.69$^\ast$ &  AFGL\,2298$^\ast$ & P\,Cyg$^\ast$ \\
\hline
\end{tabular}
\end{table*}

The census of Galactic confirmed and cLBVs presented in Clark et
al. (2005) lists 12 and 23 stars, respectively. About 60 per cent
of these stars are associated with compact circular or bipolar
circumstellar nebulae. Table\,\ref{tab:LBV} updates the list of
the bona fide LBVs. (A current census of the cLBVs will be
presented elsewhere.) The stars are arranged according to their
right ascension, which increases in the table from left to right
and from top to bottom. The names of three bona fide LBVs listed
in Clark et al. (2005) cannot be recognized by the SIMBAD data
base. For these stars we give their recognizable names and, for
convenience, in parentheses we give their names from Clark et al.
(2005).

Table\,\ref{tab:LBV} contains four new confirmed LBVs (marked by
bold face). MWC\,930 was suggested as a cLBV quite long ago
(Miroshnichenko et al. 2005) and its LBV status was confirmed this
year (Miroshnichenko et al. 2014). Discovery of a circular shell
around MWC\,930 (Cerrigone et al. 2014) lends an additional
support to the LBV status of this star.
[MMC2010]\,LBV\,G0.120-0.048 was discovered by Mauerhan et al.
(2010) in the spectroscopic follow-up of unidentified point
sources of Paschen-$\alpha$ (P$\alpha$) line excess in the
Galactic Centre. The LBV status of this star was supported by its
LBV-like spectrum, the already known photometric variability
($\approx$1 mag during several years) and the presence of a
circular nebula of P$\alpha$ emission centred on the star. Two
other new LBVs, Wray\,16-137 (Gvaramadze et al. 2014) and WS1,
were first identified as cLBVs through the detection of circular
shells and follow-up spectroscopy of their central stars, and then
confirmed as LBVs thanks to the detection of major changes in
their brightness and spectral appearance.

It is worthy to note that all four new bona fide LBVs are
surrounded by circular shells, which raises the percentage of
these stars associated with circumstellar nebulae to $\approx$70
per cent.

\section{Acknowledgements}
Spectral observations reported in this paper were obtained with
the Southern African Large Telescope (SALT). AYK acknowledges
support from the National Research Foundation (NRF) of South
Africa. LNB acknowledges the Russian Science Foundation grant
14-22-00041.

\end{document}